\documentclass[aps,prl,twocolumn,groupedaddress,amsmath,amssymb,nofootinbib]{revtex4-1}

\usepackage[utf8]{inputenc}
\usepackage{graphicx}
\usepackage{caption}
\usepackage{amsmath,amssymb}
\usepackage[colorlinks=true,allcolors=blue]{hyperref}

\newcommand{\D}{\mathrm{d}}
\newcommand{\vecr}{\boldsymbol{r}}
\newcommand{\vecq}{\boldsymbol{q}}
\newcommand{\insertpdfpage}[2]{%
  \thispagestyle{empty}%
  \noindent\includegraphics[page=#2,width=\textwidth,height=\textheight,keepaspectratio]{#1}%
  \clearpage
}

\begin{document}

\title{Collective dynamics of trail-interacting particles}

\author{Paul Pineau}
\email{paul.pineau@sorbonne-universite.fr}
\affiliation{Laboratoire Jean Perrin, CNRS/Sorbonne Université, Paris, France}

\author{Samuel Bell}

\affiliation{Laboratoire Jean Perrin, CNRS/Sorbonne Université, Paris, France}

\author{Raphaël Voituriez}
\email{raphael.voituriez@sorbonne-universite.fr}
\affiliation{Laboratoire Jean Perrin, CNRS/Sorbonne Université, Paris, France}

\author{Ram M. Adar}
\email{radar@technion.ac.il}
\affiliation{Department of Physics, Technion – Israel Institute of Technology, Haifa 32000, Israel}

\begin{abstract}
Trail interactions occur when past particle trajectories bias future motion, rendering the system out of thermodynamic equilibrium. While such systems are abundant in nature, their understanding is limited to the single-particle level or phenomenological mean-field theories. Here, we introduce a minimal model of many trail-interacting particles, which extends this paradigm to the fluctuating collective level in arbitrary dimension. Particles diffuse while depositing long-lasting repelling/attracting trails that act as a shared memory field, coupling their dynamics across time and space. Using stochastic density functional theory, we derive fluctuating hydrodynamic equations and analyze analytically and numerically the resulting behaviors. We show that memory, coupled with fluctuations, fundamentally reshapes collective dynamics; In the repulsive case, the particle density displays superdiffusive spreading for $d=1$ and diffusive spreading for $d\geq2$, characterized by transient clustering and ballistic motion; In the attractive case, the system condensates in finite time into frozen, localized states. Our results establish general principles for trail-interacting systems and reveal how persistent fields generate novel instabilities and drive self-organization.
\end{abstract}

\maketitle
  
Many living and synthetic agents modify their environment by depositing persistent, non diffusive trails that bias their subsequent motion. Examples include ants laying pheromone traces \cite{Dussutour2004, Jackson2006}, cells navigating via self-secreted cues \cite{Flyvbjerg2021, dAlessandro2021, Troger2024}, pedestrians leaving footprints on the ground~\cite{Helbing1997}, robots pushing away obstacles~\cite{Altshuler2024}, and self-propelled droplets guided by trails of micelles~\cite{Jin2017,Hokmabad2022, Chen2025}. In all these examples, the deposited trails neither degrade nor diffuse appreciably over experimental timescales, thereby providing a shared memory of past motion that drives these systems out of equilibrium, and drives  spatial and temporal correlations. The trail persistence and absence of diffusion stands in contrast to the broad class of models of autochemotactic biological systems~\cite{KellerSegel1,KellerSegel2,KellerSegel3,Brenner1998} or autophoretic particles \cite{Meyer2023,Saha2014,Romano2026}, where the deposited chemical field diffuses and memory is lost on finite timescales. Despite the ubiquity of trail-interacting systems, their collective behavior remains poorly understood.

Inspired by the variety of experimental systems mentioned above, theoretical studies of trail depositing particles have flourished in both mathematics and physics, but so far have mostly focused on self-interacting random walks, where a single particle’s trajectory is influenced by its own past \cite{Amit1983,toth95, Pemantle2007, Horvath2011, Dumaz2013, BarbierChebbah2022, Bremont2024, Bremont2025}. Such non-Markovian dynamics result in rich phenomena, including localization, anomalous diffusion, and aging. By contrast, far less is known about the collective behavior of particles coupled through the trails they all generate. As the dynamics depends on the full system's history, the collective description cannot be reduced to instantaneous pairwise interactions. Existing approaches rely mainly on mean-field or hydrodynamic descriptions \cite{Toth2002, TothValko2005, Adar2024, Hartman2024, Bell2025, Maggs2026},
as well as numerical simulations \cite{Gelimson2016, Meyer2023, Wysocki2026, Mokhtari2022, Hartman2024, Maggs2026}. 

\begin{figure}[ht]
\includegraphics[width=\columnwidth]{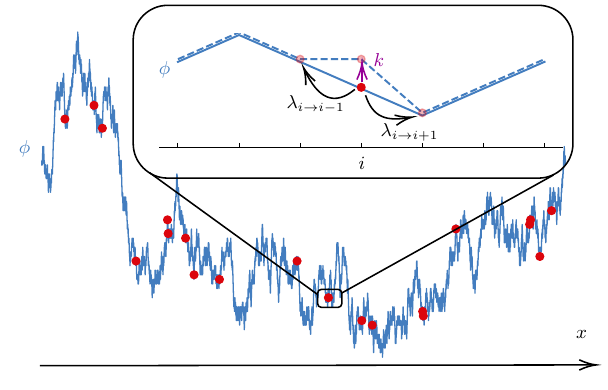}
\caption{Schematic illustration of the model in one dimension. Each particle (in red) lives in a potential landscape $U=h\phi$. Inset: two possible dynamics of a single particle: hopping to neighboring sites with rates $\lambda_{i\pm 1}$ and trail deposition with rate $k$. One deposition increase locally $\phi$ by 1 (purple event). Translucent red dots represents the three possible future state of the particle. }
\label{fig1}
\end{figure}

In this Letter, we introduce and analyze a minimal model of trail-interacting particles that extends single-particle self-interacting walks to the fluctuating hydrodynamic level in arbitrary dimension. The model couples the motion of diffusing particles to a non-decaying memory field that records their trajectories and biases subsequent motion. Using stochastic density functional theory and large-scale agent-based simulations, we show how trail interactions reshape collective dynamics, leading to nontrivial scaling regimes, propagating modes, and condensation phenomena. Our framework establishes general principles for the study of systems with long-lived self-generated fields and opens new directions for exploring memory-driven self-organization in statistical physics.

{\it Model.} We formulate a lattice model of trail-interacting particles in arbitrary dimension $d$, sketched in Fig.~\ref{fig1} for $d=1$ for simplicity.  We consider $N$ particles distributed on $(L/a)^d$ lattice sites, where $L$ is the linear system size and $a$ the lattice constant. The state of each site $i$ is described by its particle occupancy $\rho_i$ and trail content $\phi_i$. The coupled dynamics of both fields, defined for convenience in continuous time $t$, is driven by two types of independent transition events: (i) hopping of a particle from site $i$ to  a neighboring site $j$, with rate $\rho_i\lambda_{i\to j} = 2 \rho_i k_{tr}/\left(1+\exp\left[U(\phi_{j})-U(\phi_i)\right]\right)$ and (ii) deposition of a trail unit, with rate $\rho_i k $. Here, $k_{tr}$ is the single particle translation rate, related to the diffusion constant $D$ according to $D=k_{tr}a^2,$ and $U(\phi_i)$ acts as a potential that biases the hopping rate according to differences in trail concentration in neighboring sites. Describing the bias via an effective potential is common in the modeling of self-interacting random walks~\cite{BarbierChebbah2022}. For a stationary external potential, the chosen hopping rates would satisfy local detailed balance. Finally, $k$ is the trail deposition rate per particle. Note that all particles contribute equally to the trail field, such that each particle interacts with its own trail in the same way as it does with trails deposited by other particles. Importantly,  the trail is not degraded in this model,  and most importantly does not diffuse ; It thus acts as an infinitely  persistent memory field for particle trajectories, in deep contrast to the broad class of models of autochemotactic biological systems~\cite{KellerSegel1,KellerSegel2,KellerSegel3,Brenner1998} or autophoretic particles  \cite{Meyer2023,Saha2014,Romano2026}. Furthermore, unlike the classical Keller-Segel theory~\cite{KellerSegel1,KellerSegel2,KellerSegel3}, our theory accounts for fluctuations beyond MF. These fluctuations, combined with trail persistence, lead to a wide range of interesting features. The model can be generalized to additional trail dynamics (such as degradation or deterministic deposition) in a straightforward manner.

The lattice dynamics can be mapped to a stochastic density functional theory (SDFT) in the continuum limit, using exact path-integral methods \cite{Biroli1,Biroli2,Tridib1,Tridib2}. The resulting equations in arbitrary dimension $d$ are~\cite{SM}
\begin{align}
\label{eq1}
\partial_{t}\rho & =D\boldsymbol{\nabla}^2\rho+D\boldsymbol{\nabla}\left[\rho \boldsymbol{\nabla}U\left(\phi\right)\right]+\boldsymbol{\nabla}\left(\sqrt{2D\rho}\,\boldsymbol{\eta}\right),\nonumber \\
\partial_{t}\phi & =k\rho+\sqrt{k\rho}\,\xi.
\end{align}
These are the central equations of this paper. The stochastic nature of the dynamics is manifested in the Gaussian white noise terms in the particle and trail dynamics, $\boldsymbol{\eta}$ and $\xi$, respectively. The noise terms satisfy $\langle y_i\rangle=0$ and $\langle y_i(t,\vecr)y_j(t',\vecr')\rangle=\delta_{ij}\,\delta(t-t')\delta(\vecr-\vecr')$, where $y_i$ is either $\xi$ or a cartesian component of $\boldsymbol{\eta}$.
Importantly, the mapping between the lattice dynamics and Eq.~\eqref{eq1} is exact in the $a\to0$ limit, except for the $\xi$ term that approximates the true shot noise of the Poissonian deposition process. This approximation becomes exact at long times.

While trail-mediated interactions are non-Markovian and lie beyond the classical framework of pairwise interactions, their resulting dynamics have a transparent physical interpretation. According to Eq.~\eqref{eq1}, the particle density obeys a Dean-Kawasaki equation \cite{DeanLangevin,Kawasaki1998}. The particles undergo diffusion and advection with the effective potential $U\left(\phi\right)$ with multiplicative noise encoding particle number fluctuations. This apparent simplicity -- with no explicit interaction term -- comes at the cost of a coupling to the additional trail density field $\phi$, 
which undergoes a simple irreversible first-order growth kinetics with source term $\rho$ and multiplicative noise originating from the stochastic deposition. We focus below on the linear potential $U=h\phi$. In the single-particle case, it corresponds to the class ``True self-avoiding walks'' (TSAW). As is evident below, this simple linear form suffices to drive a wide range of effects. We analyze the linear stability of the system within mean field (MF) and proceed by analyzing correlations beyond MF in both the repulsive and attractive cases ($h>0$ and $h<0$, respectively). 

 {\it Linear stability analysis and time dependent structure factors.} We consider first a simple uniform state with $\rho(t=0,\boldsymbol{r})=\rho_0$ and $\phi(t=0,\boldsymbol{r})=\phi_0$. Without loss of generality, we set $\phi_0=0$. We linearize the equations around the uniform dynamics by introducing the first-order corrections in the fields $\delta\rho(t,\boldsymbol{r})=\mathrm{Re} \left[\delta\rho(t,\boldsymbol{q})\exp \left(i\boldsymbol{q\cdot \vecr}\right)\right]$ and similarly for $\phi$, so that $\rho=\rho_0+\delta\rho$ and $\phi=k\rho_0 t+\delta\phi$. Furthermore, we make use of   dimensionless forms, where lengths are normalized by the system size $L$, times by $\tau=L^2/D$, and densities by the average density $\rho_0$. For simplicity, we use the same notations, such that $\rho\to\rho_0\rho,$ $\phi\to\rho_0\phi,$ $\partial_t\to\partial_t/\tau$, $\boldsymbol{q}\to \boldsymbol{q}/L$, $k\to k/\tau$, and $h\to h/\rho_0$. This leads to 
\begin{align}
\label{eq2}
\partial_{tt}\delta\rho+q^2\partial_t\delta\rho+khq^2\delta\rho=\frac{\Xi}{\sqrt{N}},
\end{align}
with the noise term $\Xi=\sqrt{2}i \boldsymbol{q}\cdot \partial_t \boldsymbol{\eta}-\sqrt{k}hq^2\xi$ and $N = \rho_0 L^d$ the total number of particles. As the system is isotropic, it only depends on the modulus $q=|\vecq|$. This is the equation of a damped harmonic oscillator, where $q^2$ is the friction coefficient, originating in diffusion, while $khq^2$ plays the role of the spring constant, originating in the memory field. It is clear that the system is linearly unstable for all $q$ for an attractive interaction $(h<0),$ corresponding to a negative spring constant. Explicitly, upon Fourier transformation in the time domain, the dispersion relation is given by $i\omega=-q^2/2\pm\sqrt{(q^2/2)^2-khq^2}$.

To go beyond the linear stability analysis, we make use of our exact determination of the noise amplitudes and predict  the dynamics of the static structure factors of the dimensionless particles density $S_\rho(t,q)=\langle \delta\rho(t,q)\delta\rho(t,-q)\rangle$ and dimensionless trail density $S_\phi(t,q)=\langle \delta\phi(t,q)\delta\phi(t,-q)\rangle$. Eq.~\eqref{eq2} is solved and averaged over noise realizations, leading to~\cite{SM} 
\begin{align}
\label{eq3}
S_\rho(t,q)&=\frac{q^2}{N}\int_{0}^{t}ds\left[2G'^2(s,q)+kh^2q^2G^2(s,q)\right],\\
S_\phi(t,q)&=\frac{k}{N}\int_{0}^{t}ds\left[\left(G'(s,q)+q^2G(s,q)\right)^2 + 2kq^2G^2(s,q)\right], \nonumber
\end{align}
where the kernel is defined as
\begin{align}
\label{eq4}
G(\tau,q)&=\frac{e^{-q^2 \tau/2}}{\sqrt{\left(q^2/2\right)^2-q^2 kh}}\sinh\left(\tau \sqrt{\left(q^2/2\right)^2-q^2 kh}\right),
\end{align}
and $G'=\partial G/\partial\tau$. These results rely on the linearized dynamics and are valid for sufficiently small noise amplitudes. The explicit time dependence that we find  captures the transient dynamics from the uniform initial condition, which is a stationary state of the MF dynamics of the particle density field, but not of the full fluctuating dynamics. In what follows, we analyze the results of Eqs.~\eqref{eq3}-\eqref{eq4}, while comparing analytical predictions to lattice simulation results for $d=1,2$ (see SM \cite{SM} and \cite{CODE_collective}). We consider the repulsive (linearly stable) and attractive (linearly unstable) cases separately, as they display qualitatively different behaviors.

\begin{figure}[ht]
\centering
    \includegraphics[width=\columnwidth]{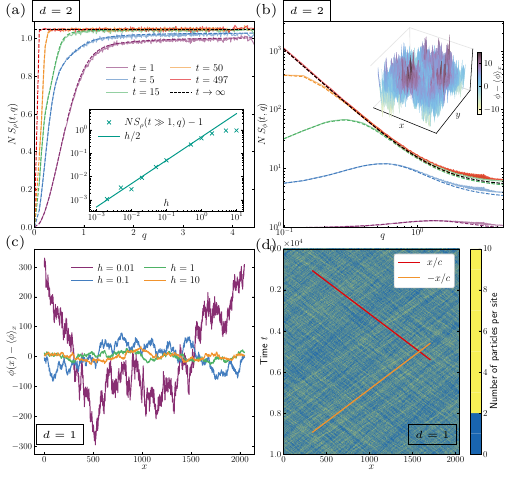}
\caption{Repulsive case ($h>0$). (a+b) $d=2$: Density- and trail structure factors as a function of the wavevector at different times. Simulation results are plotted as solid lines and the theoretical prediction of Eq.~\eqref{eq3} as dashed lines. The asymptotic value for long times is plotted in dashed black. Inset of (a) shows how the long-time density structure factor scales with $h$, according to the simulations (symbols) and theory (solid line). Inset of (b) shows a realization of the trail field $\phi$ at steady-state for $h=0.1$ (c) Characteristic trail profiles in $1d$ at long times for different $h$ values, according to the simulations. (d) Kymograph of density in $1d$ as a function of space and time. The texture indicates that low- and high-density fluctuations travel at a velocity $\pm c=\pm\sqrt{hk}$. 
}
\label{fig2}
\end{figure}

{\it Repulsive case $(h>0)$. } 
In the repulsive case, the uniform state is linearly stable as  shown above. Nevertheless, both particles and trail structure factors display a non-trivial steady state and transient dynamics, which are captured remarkably well, with no adjustable parameter, by our linear analysis  for both $d=1$ and $d=2$ (see Fig.~\ref{fig2}a,b). At long times  and for a non vanishing $h>0,$ the trail structure factor is found to reach a steady state (see Fig.~\ref{fig2}b,c)
\begin{equation}
\label{eq5}
NS_\phi(t,q)\underset{t\to\infty}{\approx}\frac{k(2+h)+q^2}{2hq^2}.
\end{equation}
This result interpolates between two simple limits: for small wavevectors, $q\ll\sqrt{k(2+h)}$, the structure factor scales as $\sim1/q^2,$ which is consistent with a trail profile $\phi(x)$ that behaves as an equilibrium Edwards-Wilkinson interface  in quantitative agreement with simulations for both $d=1$ and $d=2$ (see Fig.~\ref{fig2}b,c and~\cite{SM}). Note that in one dimension, such a steady state profile is also consistent with that of the Kardar-Parisi-Zhang equation~\cite{Takeuchi2018,Hairer2018}. Interestingly, the trail field was found earlier to be exactly Edwards-Wilkinson in the case of a single true self-avoiding walker in infinite space with comparable trail mediated dynamics for $d=1$ \cite{toth95}. A non--linear analysis, necessary to fully characterize the trail field, is left for further work. For large wavevectors, $q\gg\sqrt{k(2+h)}$, the structure factor scales as $\sim q^0,$ due to the noise in the deposition process itself, which induces Poissonian fluctuations. Of note, the $h\to 0$ limit is singular in Eq.~\eqref{eq5} and indicates that a non-vanishing interaction with the trail field is necessary for the trail fluctuations to reach a steady state. The case $h=0$ is discussed in~\cite{SM}.

For $h>0$, the particle static structure factor also reaches a steady state, which is found analytically (and verified numerically for $d=1,2$ in Fig.~\ref{fig2}a and SM \cite{SM}) to be independent of $q$:
\begin{equation}
\label{eq5bis}
N \,S_\rho(t,q)\underset{t\to\infty}{\approx}1+h/2.
\end{equation}
This indicates that, counterintuitively, {\it repulsive} trail-mediated interactions lead at steady state  to effective zero-range {\it attractive} interactions. The static structure factor is that of (on average) $N/(1+h/2)$ independent one-lattice-site clusters of $1+h/2$ particles. 
To decipher the microscopic origin of such zero-range attractive interactions, we analyse  
 the dynamic structure factor $S_\rho(\omega,q)=\langle| \delta\rho(\omega,q)|^2 \rangle$, which  can be obtained for all $d$ upon  Fourier transformation in time domain of Eq.~\eqref{eq2}. This yields~\cite{SM}
\begin{equation}
\label{eq6}
NS_\rho(\omega,q)= \frac{1}{q^2} \frac{2 + \frac{h c^2 q^2}{\omega^2}}{1 + \frac{\omega^2}{q^4}\left(1 - \frac{c^2q^2}{\omega^2}\right)^2}\equiv\frac{1}{ q^2}g\left(\frac{\omega}{q^2},\frac{\omega}{cq}\right)
\end{equation}
where $c=\sqrt{hk}$ is a typical velocity and $g(u,v)$ the scaling function. The dynamic structure factor of the trail field is obtained along the same method and is given in \cite{SM}, where the quantitative agreement with numerical simulations for $d=1,2$ is shown. For $q\gg  c$, the structure factor is found to satisfy a single diffusive scaling  $S_\rho(\omega,q)\sim g(\omega/q^2, v\to\infty)/q^2$, identical to that of non-interacting diffusing particles. In turn, for $q\ll c $ we find a ballistic scaling $S_\rho(\omega,q)\sim g(u\to\infty,\omega/(cq))/q^2$ with a pole for $\omega, q$ satisfying the dispersion relation $\omega=\pm cq$. This reveals the emergence of traveling waves of velocity $c$ induced by trail-mediated interactions at large length scales, which are indeed observed in numerical simulations with quantitative agreement for $d=1$ (see Fig.~\ref{fig2}d). 

Such traveling waves share similarities with wave patterns observed for repulsive autophoretic particles \cite{Meyer2023} and solutions of the Fisher-KPP equation~\cite{vanSaarloos}. It can also be understood in terms of a non-reciprocal interaction, where the trail field is ``attracted'' to the particles who, in turn, are repelled by the trail.  Furthermore, these waves provide a microscopic mechanism to account for the effective zero-range attractive interaction observed above: particles in the same wave packet experience the same local trail profile and become correlated. We note that the MF version of Eq.~\eqref{eq1} further allows for traveling-front solutions, whose profile can be found analytically from the nonlinear equations \cite{SM}.

To validate the applicability of our hydrodynamic equations Eq. (\ref{eq1}) beyond linear order, we  next analyze within MF the spreading dynamics of a localized initial condition, $\rho(t=0,\vecr)=N\delta(\vecr)$ and $\phi(0,\vecr)=0$, keeping $h>0$. We look for self-similar isotropic solutions of the form 
$\rho(t,r) = t^{-\alpha}r^{1-d}f\left(r/t^{\alpha}\right)$, 
$\phi(t,r) = t^{1-\alpha} r^{1-d}F\left(r /t^{\alpha}\right)$ in the limit $r =|\vecr|,t\to\infty$ with $u = r/t^\alpha$ finite. For $d=1$ we find a superdiffusive scaling $\alpha=2/3$ as well as the exact profiles of the particle and trail fields (see SM~\cite{SM}), in agreement with the scaling limit of Ref.~\cite{Maggs2026}.

\begin{figure}[ht]
    \centering
    \includegraphics[width=\columnwidth]{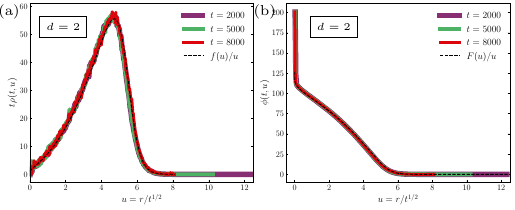}
    \caption{Rescaled profiles of (a) the particle density and (b) the trail density for $d=2$, compared with the theoretical predictions of Eq.~\eqref{eq:scaling_solution_2D}. Simulations correspond to $N=4096$, $h=0.1$, $k=1$, averaged over 100 realizations.}
    \label{fig:scalings_2d}
\end{figure}

For $d=2$, substituting the scaling solutions in the dimensionless form of Eq. (\ref{eq1}) while setting $\boldsymbol\eta,\xi=0$, yields
\begin{align}
    -\alpha (fu)^\prime &=  t^{-2\alpha+1} \left(\left(f/u\right)^\prime u\right)^\prime 
    + h t^{2 - 4\alpha} \left(f \, \left(F/u\right)^\prime \right)^\prime \nonumber\\
    kf&=(1-\alpha)F-\alpha F^\prime u,\label{eq8}
\end{align}
where $y'=\D y/\D u$. Self-similarity imposes a single diffusive scaling $\alpha = 1/2$; the coupled non-linear equations for $f(u)$ and $F(u)$ can then be solved analytically (see \cite{SM}) and yield
\begin{align}
    \frac{f(u)}{u} &= \frac{e^{-\frac{u^2}{4}}}{C - h k \text{Ei}\left(-\frac{u^2}{4}\right)}\nonumber\\
    \frac{F(u)}{u} &=  \frac{1}{h} \ln\left(1 -\frac{h}{C} \text{Ei}\left(-\frac{u^2}{4}\right)\right)\label{eq:scaling_solution_2D}
\end{align}
where Ei is the exponential integral function and $C$ a constant enforcing $\int_0^\infty\rho(t,r)2\pi r \,dr = N$. Numerical simulations confirm quantitatively this scaling form (Fig. \ref{fig:scalings_2d}) \cite{Args_Log_corrections}. Despite the diffusive scaling, it interestingly displays a non-Gaussian, non-monotonic behavior induced by the repulsive interaction with the trail field. We emphasize that this predicted density profile is exact in the MF limit, which here amounts to taking $N\to\infty$. In particular, it differs quantitatively  from that of the distribution function of a single trail-interacting particle~\cite{Maggs2026}. Here, each particle interacts equivalently with the accumulated trails of all particles, and not only with its own trail, so that relative fluctuations of the trail field become irrelevant in the large $N$ limit, while they remain important for $N=1$. Last, for $d\geq3$ the scaling analysis indicates that the interaction term is irrelevant at long times~\cite{SM} and Gaussian diffusion dominates, as expected from the simple diffusive dynamics of single particles \cite{Horvath2011,BarbierChebbah2022}.

\begin{figure}[ht]
    \includegraphics[width=\columnwidth]{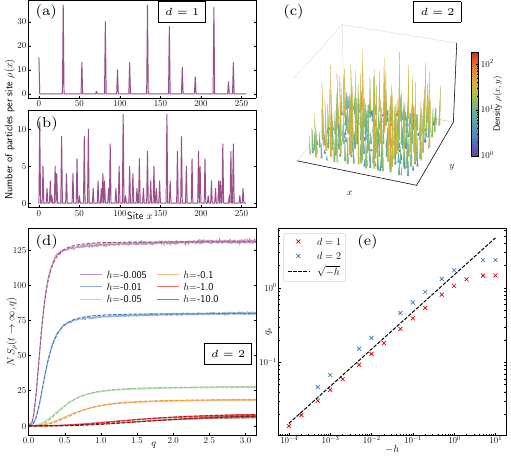}
\caption{Attractive case ($h<0$). (a-c). Snapshots of the density profile at steady state for (a) $d=1, h = 0.05, k=0.1$, (b) $d=1, h=0.1, k=1$, and (c) $d=2, h=0.01, k=1$. (d) Density structure factor as a function of the wavevector at steady-state for different $h$ values (solid lines) and their sigmoid-type fit (dashed lines). (e) The characteristic wavevector $q_\ast$ as a function of $h$ for $k=1$. The theoretical scaling is plotted in dashed black.}
\label{fig3}
\end{figure}

{\it Attractive case $(h<0)$.} The attractive case is linearly unstable, as shown above. Therefore, the analytical results obtained by analyzing perturbations around the uniform steady state are expected to hold only at short times, as is verified quantitatively by our simulations~\cite{SM}.  At longer times, we find numerically for $d=1,2$ that particles condensate in well-separated sites and remain frozen (Fig.~\ref{fig3}a,b,c), generalizing earlier results showing the self-trapping of single particles \cite{BarbierChebbah2022}. 
Qualitatively, once a condensate starts to form, the trail field at that site builds up, creating an increasingly deep potential well from which the particles cannot escape. 
While the system is far from the initial uniform steady state, the density structure factor can be heuristically understood based on the wavevector $q_\ast$ that characterizes the spatial distribution of condensates. We determine $q_\ast$  by fitting the simulation results (Fig.~\ref{fig3}d) with a sigmoid-type ansatz $S_\rho\sim \, q^n/\left(q^n+q_\ast^n\right)$, where $n=3$ is our ad-hoc choice. For $q\gg q_\ast$, the structure factor saturates to a value denoted as $S_p$, determined by the distribution of particles in each condensation site, indicating the absence of interactions between condensates at scales $\lesssim1/q_\ast$. For small $q\ll q_\ast,$  the structure factor vanishes, as the dynamics are frozen before fluctuations can reach length scales  larger than $1/q_\ast$. The freezing of the dynamics -- and thus arrested coarsening -- beyond the scale $1/q_\ast$ is due to the unbound form of the potential and the infinite persistence and non-diffusivity of the trail field; it stands in stark contrast to the condensation dynamics of the class of Keller-Segel models \cite{KellerSegel1,KellerSegel2,KellerSegel3,Brenner1998}.

The steady state in the attractive case is thus fully characterized by $q_\ast$ and $S_p$. The scaling of $q_\ast$ with $h$  can be deduced by comparing the timescales of diffusion at the scale $1/q_\ast$,  and of growth of the trail-mediated interaction $h\phi$ to significantly impact advection. For weak interactions, equating the rates $q_\ast^2$ for diffusion and $k|h|$ for the growth of the trail-mediated interaction, we find $q_\ast\sim\sqrt{k|h|}$, as validated by  simulations for $d=1,2$ (Fig.~\ref{fig3}e). Note that $|h|\ll1$ corresponds to a single condensate in finite systems, similar to a zero-range process~\cite{Evans2000}.  For strong interactions, however, a single deposition is enough to induce condensation. We thus estimate the rate of a single deposition by any of the particles found along  a distance $\sim1/q_\ast.$ This rate scales as $k/q_\ast^d$ and, equating to the diffusion rate, we find $q_\ast\sim k^{1/(d+2)}$, as validated by simulations for $d=1,2$ \cite{SM}. 

{\it Discussion.} We have introduced a minimal model for trail-interacting particles in arbitrary space dimension $d$ that extends the physics of self-interacting random walks from single-particle to collective dynamics and goes beyond available mean-field hydrodynamic descriptions. By combining stochastic density functional theory and simulations, we demonstrated how the coupling to an infinitely persistent, non-diffusive trail field that encodes the memory of the system, together with fluctuations beyond mean field, lead to clustering, propagating modes, superdiffusion, and condensation. Finite trail persistence, arising from evaporation or degradation, is expected to modify the long-time behavior, including the rough trail landscape in the repulsive case and the frozen condensates in the attractive case. We nevertheless expect the propagating modes in the repulsive case and the tendency toward phase separation in the attractive case to remain robust. Quantifying how finite persistence affects these regimes is left for future work. 

Our results highlight trail mediated memory as a generic mechanism for self-organization, distinct from direct interactions or external fields. Beyond providing a unifying framework for trail-mediated processes in biological and synthetic systems, our study suggests new directions for exploring memory-driven phenomena in nonequilibrium statistical physics, from collective migration and chemotaxis to artificial active materials.

{\it Acknowledgements.} We thank J. Brémont, O.  Bénichou, M. Aksil and Y. Kafri for useful discussions. Support from ERC synergy grant SHAPINCELLFATE and Israel Science Foundation (ISF) under Grant No. 444/25 is acknowledged.

\clearpage
\onecolumngrid
\pagestyle{empty}
\insertpdfpage{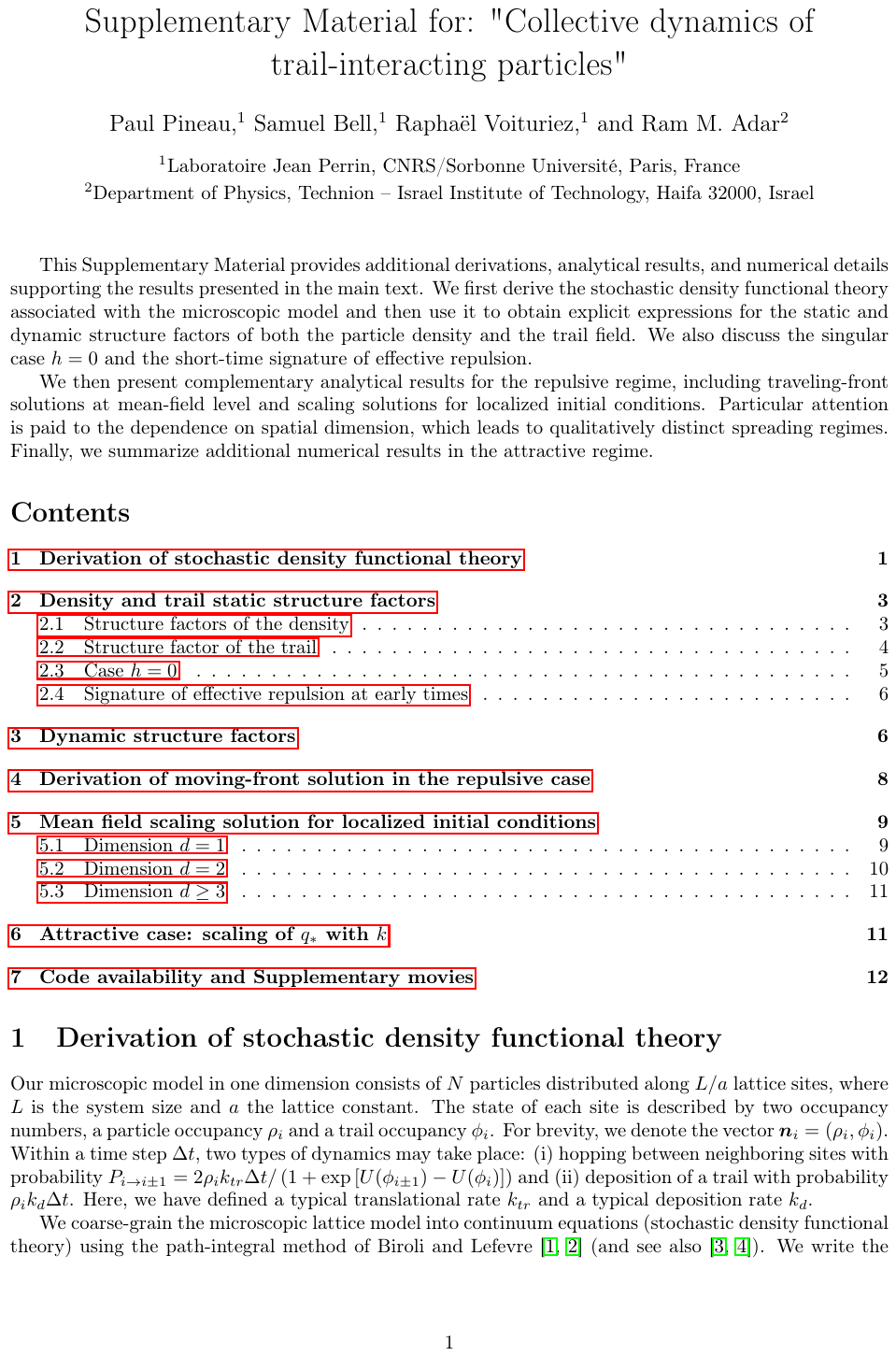}{1}
\insertpdfpage{Supplementary.pdf}{2}
\insertpdfpage{Supplementary.pdf}{3}
\insertpdfpage{Supplementary.pdf}{4}
\insertpdfpage{Supplementary.pdf}{5}
\insertpdfpage{Supplementary.pdf}{6}
\insertpdfpage{Supplementary.pdf}{7}
\insertpdfpage{Supplementary.pdf}{8}
\insertpdfpage{Supplementary.pdf}{9}
\insertpdfpage{Supplementary.pdf}{10}
\insertpdfpage{Supplementary.pdf}{11}
\insertpdfpage{Supplementary.pdf}{12}

\end{document}